\documentclass[reprint,superscriptaddress, amsmath,amssymb, aps, pra, longbibliography]{revtex4-1}

\usepackage{soul}
\usepackage{amsmath}   
\usepackage{amssymb}
\usepackage{mathtools}
\usepackage{graphicx}
\usepackage{dcolumn}
\usepackage{bm}
\usepackage{amsmath}
\usepackage{graphicx}
\usepackage{amsfonts}
\usepackage{subfigure}
\usepackage{graphicx}
\usepackage{array}
\usepackage{float}
\usepackage{color}
\usepackage{multirow}
\usepackage{amssymb}%
\usepackage[colorlinks=true,linkcolor=blue]{hyperref}%
\hypersetup{allcolors=blue}
\usepackage[normalem]{ulem}
\usepackage{xcolor}

\usepackage{mathtools}
\DeclarePairedDelimiterX\braket[2]{\langle}{\rangle}{#1 \delimsize\vert #2}

\begin{document}
\title{High-angular-momentum Rydberg states in a room-temperature vapor cell for DC electric-field sensing}
\date{\today }

\author{Alisher Duspayev}
    \email{alisherd@umich.edu}
    \affiliation{Department of Physics, University of Michigan, Ann Arbor, MI 48109, USA}
\author{Ryan Cardman}
    \thanks{Present address: Physical Sciences Inc., Andover, MA 01810, USA}
    \affiliation{Department of Physics, University of Michigan, Ann Arbor, MI 48109, USA}
\author{David A. Anderson}
    \affiliation{Rydberg Technologies Inc., Ann Arbor, MI 48103, USA}    
\author{Georg Raithel}
    \affiliation{Department of Physics, University of Michigan, Ann Arbor, MI 48109, USA}  

\begin{abstract}
We prepare and analyze Rydberg states with orbital quantum numbers $\ell \le 6$ using three-optical-photon electromagnetically-induced transparency (EIT) and radio-frequency (RF) dressing, and employ the high-$\ell$ states in electric-field sensing. Rubidium-85 atoms in a room-temperature vapor cell are first promoted into the $25F_{5/2}$ state via Rydberg-EIT with three infrared laser beams. Two RF dressing fields then (near-)resonantly couple $25 \ell$ Rydberg states with high $\ell$. The dependence of the RF-dressed Rydberg-state level structure on RF powers, RF and laser frequencies is characterized using EIT.    
Furthermore, we discuss the principles of DC-electric-field sensing using high-$\ell$ Rydberg states, and experimentally demonstrate the method using test electric fields of $\lesssim$~50~V/m induced via photo-illumination of the vapor-cell wall. We measure the highly nonlinear dependence of the DC-electric-field strength on the power of the photo-illumination laser. Numerical calculations, which reproduce our experimental observations well, elucidate the underlying physics. Our study is relevant to high-precision spectroscopy of high-$\ell$ Rydberg states, Rydberg-atom-based electric-field sensing, and plasma electric-field diagnostics.
\end{abstract}

\maketitle

\section{Introduction}
\label{sec:intro}

Rydberg-atom-based sensors for electromagnetic fields  are a prominent example how atomic-physics research is being translated 
into applications (for recent reviews see, for example,~\cite{Adams_2020, Anderson.RydergCommsSensing.2020, fancher2021, 9748947, Yuan_2023}), offering high sensitivities, large bandwidth, traceability and self-calibration~\cite{Holloway2014, RFMS}.
Modern Rydberg field sensors avoid complex apparatuses by the use of room-temperature vapor cells and electromagnetically-induced transparency (EIT)~\cite{boller1991} for optical interrogation of field-sensitive Rydberg atoms~\cite{mohapatra2007}, allowing hybridization with existing classical technologies and the realization of portable instrumentation~\cite{Simons:18,andersonapl2018,RFMS,  meyer2021,mao2023}. Beyond demonstrations of electric field sensing with atomic Rydberg states~\cite{osterwalder, Sedlacek2012, Holloway2014}, Rydberg sensors have proven their potential utility in modulated signal reception from long-wavelength radio-frequencies to millimeter-wave~\cite{anderson2021, holloway2021, Prajapati2022,legaie2023}, phase detection~\cite{anderson2022, berweger2023, 10096264}, polarization measurements~\cite{sedlacek2013,jiao2017,andersonapl2018,anderson2021,Wang:23}, spatial field mapping and RF imaging~\cite{hollowayapl2014, downes2020, CardmanAOT2020, Anderson.2023} and in the establishment of new atomic measurement standards~\cite{Sedlacek2012, sedlacek2013, powerstandard2018, voltstandard2022, gaugeeffects}.

More recently, two-photon Rydberg-EIT has been generalized to three optical photons~\cite{carr2012, Moore2019a, prajapati2023}, which afford operation at all-infrared wavelengths~\cite{thoumany2009, Fahey11, johnson2012, You:22}, allow efficient Doppler-shift cancellation conducive to narrower EIT linewidth~\cite{bohaichuk2023}, and offer inroads towards excitation of high-angular-momentum ($\ell \geq 3$) Rydberg states for electric-field detection in very-high and ultra-high frequency bands~\cite{brown2023, elgee2023, prajapati2023highell}. High-$\ell$ Rydberg states exhibit large sensitivities to DC fields~\cite{ma2022} due to their DC electric polarizabilities, which scale as $n^7$ (with  principal quantum number $n$) and also increase with $\ell$~\cite{gallagher}. Hence, $n$ and $\ell$ can be traded in favorable ways to achieve high DC-field sensitivity at a comparatively low $n$, where optical excitation pathways are more efficient than at high $n$, and where Rydberg-atom interactions (which typically scale with high powers of $n$) are less limiting. These aspects are expected to be particularly relevant to field sensing in plasmas~\cite{alexiou1995,feldbaum2002,borghesi2002,park2010,anderson2017,weller2019,Chng_2020, Goldberg_2022} and ion-beam sources~\cite{smith2006, bassim2014recent, bischoff, gierak}. For potential use in the latter~\cite{knuffman2013, claessens, McCulloch_2016}, methods for Rydberg EIT plasma field measurement and diagnostics have been developed~\cite{anderson2017} and microfield sensing using Rydberg atoms have recently been demonstrated~\cite{ionsourcepaper}.

Motivated by prospects in plasma-field sensing, here we investigate the response of high-$\ell$ Rydberg atoms to electric fields in a vapor cell. We prepare $25F$, $25G$, $25H$ and $25I$ ($\ell = 3$ to 6) Rydberg states in rubidium-85 ($^{85}$Rb) using three-optical-photon EIT and up to two radio-frequency (RF) dressing fields. The efficacy of high-$\ell$ Rydberg-state preparation is verified by measurements of Autler-Townes (AT) splittings induced by the RF dressing fields.  DC electric fields are created by photo-illumination of the dielectric vapor-cell wall using 453-nm light~\cite{ma20}, and the response of the high-$\ell$ Rydberg states to the DC fields is explored.  The paper is organized as follows. In Sec.~\ref{sec:high-ell} we outline the benefits of high-$\ell$ Rydberg states for DC-field sensing. In Sec.~\ref{sec:exc} we describe our experimental setup and the utilized RF dressing methods. In Sec.~\ref{sec:dcsensing} we explain our method of creating DC electric fields by illuminating the cell wall with a $453$-nm laser. We then present our DC-field sensing results, which include an approximate calibration of the weak, DC field in the vapor cell versus $453$-nm illumination power. In the concluding Sec.~\ref{sec:disc}, possible applications and research venues for future work are discussed.

\section{High-$\ell$ Rydberg states in static electric fields}
\label{sec:high-ell}

\begin{figure}[t!]
 \centering
  \includegraphics[width=0.45\textwidth]{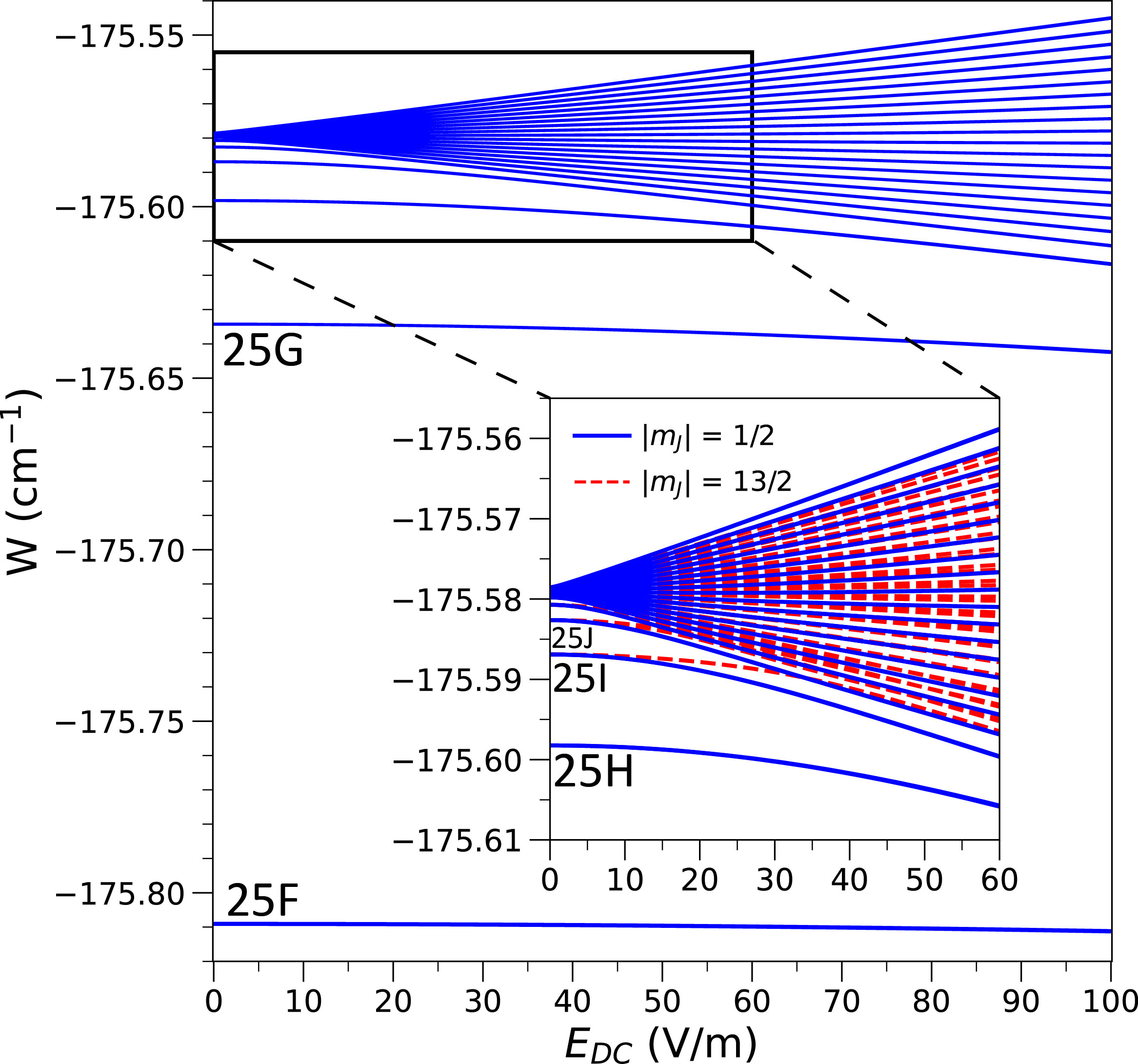}
  \caption{Computed Stark maps for $^{85}$Rb $n=25$ and two sample cases of $|m_J|$, and zoomed-in detail in the inset, illustrating the increasingly sensitive quadratic response of Rydberg atoms to weak DC electric fields as $\ell$ increases from $\ell=3$ (25$F$) to $\ell=7$ (25$J$). The fine structure is not resolved in this figure.} 
  \label{fig:Stark_t}
\end{figure}

\begin{table}[t!]
\caption{\label{tab:table1} Calculated scalar and tensor, and $|m_J|$-dependent (see Eq.~\ref{eq:alpha}) DC electric polarizabilities, $\alpha_S$, $\alpha_T$, and $\alpha_{DC}(|m_J|)$ respectively, [all in MHz/(V/m)$^2$] for the high-$\ell$ Rydberg states used in our study.}
\begin{tabular}{|c |c |c |c |c |}
    \hline
    State & $\alpha_S$ & $\alpha_T$ &$|m_J|$ & $\alpha_{DC}$\\
    \hline
    \multirow{3}{*}{$25F_{5/2}$} & \multirow{3}{*}{0.010037} & \multirow{3}{*}{-0.003492} & 1/2 & 0.012830\\
                                 & & & 3/2 & 0.010735\\
                                 & & & 5/2 & 0.006544\\
    \hline                             
    \multirow{4}{*}{$25G_{7/2}$} & \multirow{4}{*}{0.039466} & \multirow{4}{*}{-0.014841} & 1/2 & 0.050067\\
                                 & & & 3/2 & 0.045827\\
                                 & & & 5/2 & 0.037346\\
                                 & & & 7/2 & 0.024626\\
    \hline  
    \multirow{5}{*}{$25H_{9/2}$} & \multirow{5}{*}{0.106447} & \multirow{5}{*}{-0.042547} & 1/2 & 0.134812\\
                                 & & & 3/2 & 0.127720\\
                                 & & & 5/2 & 0.113538\\
                                 & & & 7/2 & 0.092264\\
                                 & & & 9/2 & 0.063899\\
    \hline  
    \multirow{6}{*}{$25I_{11/2}$}& \multirow{6}{*}{0.247529} & \multirow{6}{*}{-0.105721} & 1/2 & 0.314806\\
                                 & & & 3/2 & 0.303273\\
                                 & & & 5/2 & 0.280206\\
                                 & & & 7/2 & 0.245606\\
                                 & & & 9/2 & 0.199474\\
                                 & & & 11/2& 0.141808\\
    \hline      
\end{tabular}
\end{table}

Rydberg atoms are very sensitive to external electric fields~\cite{gallagher}. Applications discussed in  Sec.~\ref{sec:intro} include the measurement of electric fields in plasmas and ion sources.  In a recent cold-atom study~\cite{ionsourcepaper}, $\ell=1$ ($P$-) states were employed at relatively high-$n$ to measure electric fields on the order of a few tens of~V/m with $\ell = 3$ ($F$-) states to increase sensitivity to weaker electric fields. In this work, we increase the Rydberg-state electric-field sensitivity by extending to higher $\ell$ Rydberg states beyond $\ell = 3$, while also using considerably lower $n$ Rydberg states, which exhibit larger optical oscillator strengths, scaling as $n^{-3}$~\cite{gaugeeffects, gallagher}, affording stronger spectroscopic signals. Lower-$n$ states also provide the benefit of being less sensitive to Rydberg-atom interactions~\cite{Han2018, Han2019,Gaj.2014}, making them suitable for Rydberg-atom applications in inert buffer gases and in low-pressure discharge plasmas. The latter applications benefit from a room-temperature vapor-cell platform~\cite{anderson2017,ma20}, which we employ in the present work. Electric polarizabilities increase quickly with $\ell$, which is key to reducing $n$ while maintaining a desired level of field sensitivity. Lastly, electric-dipole transition matrix elements between high-$\ell$ Rydberg states are large ($\sim n^2$)~\cite{gaugeeffects, gallagher}, allowing efficient RF dressing to access higher-$\ell$ states with relatively weak RF fields.

In Fig.~\ref{fig:Stark_t} we have calculated Stark maps of Rb Rydberg states near $n=25$ for two exemplary values of $|m_J|$, the conserved component of the electronic angular momentum in the direction of the DC electric field, $E_{DC}$. 
The Stark maps visualize the quadratic Stark effect of the levels
$\vert n=25, \ell, J, m_J \rangle$, indicating a large increase in polarizability from $\ell=3$ to $7$. The quadratic Stark shifts, 
\begin{equation}
\Delta W = -\frac{1}{2} \alpha_{DC} |E_{DC}|^2,
\label{eq:DCStark}
\end{equation}
are given by state-dependent electric polarizabilities, $\alpha_{DC}$. The latter can be calculated as

\begin{equation}
\alpha_{DC} (m_J) = \alpha_S + \alpha_T \frac{3 m_J^2 - J(J+1)}{J (2J-1)},
\label{eq:alpha}
\end{equation}

\noindent with $(n, \ell, J)$-dependent scalar and tensor polarizabilities $\alpha_S$ and $\alpha_T$, respectively. Table~\ref{tab:table1} lists calculated values of $\alpha_S$, $\alpha_T$ and $\alpha_{DC}$ for the $n=25$ high-$\ell$ Rydberg states utilized in our study. As one can see, the tensor contributions are substantial. In the experiments discussed in the following Sections, only states with $|m_J| = 1/2, 3/2, 5/2$ are coupled and relevant to electric-field sensing. It is apparent that $\alpha_{DC}$ increases by a factor of $\approx$~3 with each increment in $\ell$. Thus, with our progression from $F$ to $I$ states Stark shifts increase by about a factor of 30.
Here we are able to measure Stark shifts as low as about 3~MHz, corresponding to $E_{DC}$ sensitivity limits of $\sim$20~V/m and $\sim$4~V/m for $25F$ and $25I$ states, respectively. These sensitivities are suitable for measuring electric fields in plasmas. The approximate $\sim n^{-7/2}$-scaling of the field sensitivity as a function of $n$ may be used to match sensitivity requirements elsewhere.

\section{Excitation of $\ell \le$~6 Rydberg atoms in a room-temperature vapor cell}
\label{sec:exc}

We use a multi-photon scheme with up to six photons to access high-$\ell$ states for electric-field sensing in a vapor cell. We first describe the optical three-photon Rydberg-EIT setup, which provides optical spectroscopic access to the Rb $25F_{5/2}$ Rydberg states. Then, we describe the three modes of strong RF drive using additional one-, two-, and three-RF-photon couplings, which we employ to extend the state space to include $25G_{7/2}$, $25H_{9/2}$ and $25I_{11/2}$ states, respectively. These high-$\ell$ states are then utilized to measure weak DC electric fields.

\begin{figure}[t!]
 \centering
  \includegraphics[width=0.35\textwidth]{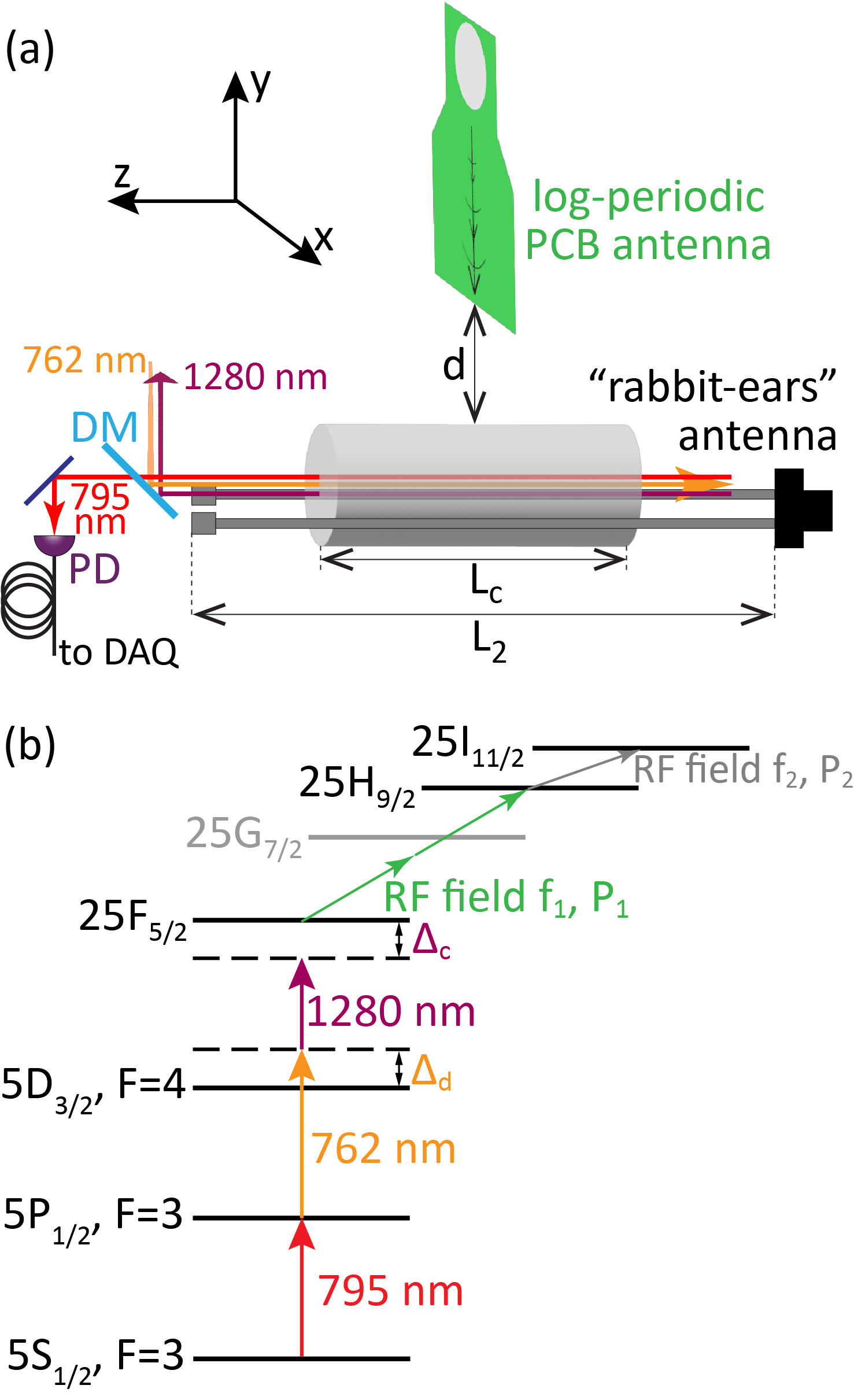}
  \caption{Sketch of the experimental setup (a) and the energy-level diagram (b) (not to scale). See text for details.}
  \label{fig:setup}
\end{figure}

\subsection{Experimental Setup}
\label{subsec:exp}

We concentrate on the $^{85}$Rb isotope, but analogous experiments could be performed with other Rb isotopes and atomic species. Three laser beams are aligned co-linearly into a room-temperature cell of length $L_c\approx 7.5$~cm filled with a natural Rb vapor, as illustrated in Fig.~\ref{fig:setup}~(a). The energy-level diagram in Fig.~\ref{fig:setup}~(b) identifies the utilized atomic transitions. The three lasers driving transitions have approximate wavelengths of 795~nm, 762~nm and 1280~nm, and are referred to as EIT probe, dressing, and coupler lasers, respectively~\cite{carr2012,Moore2019a}. The probe laser is an external-cavity diode laser (ECDL) that is locked to the Doppler-free $5S_{1/2}, F=3 \rightarrow 5P_{1/2}, F=3$ transition on the $^{85}$Rb D1 line using saturation spectroscopy. A small portion of the 795-nm light is red-shifted by $\Delta_r = -42$~MHz and counter-aligned with a small portion of the 762-nm dressing laser, which also is an ECDL, in a separate two-photon EIT reference vapor cell. The 762-nm dressing laser is locked to the $5P_{1/2}, F=3 \rightarrow 5D_{3/2}, F=4$ EIT line in this separate cell, resulting in a shift of $\Delta_d = - \Delta_r \frac{795~{\rm{nm}}}{762~{\rm{nm}}} = 44$~MHz. In the measurement cell in Fig.~\ref{fig:setup}~(a), the 795-nm laser probes the $5S_{1/2}, F=3 \rightarrow 5P_{1/2}, F=3$ resonance for atoms near zero velocity along the $z$-axis, and the 762-nm laser is blue-detuned by $\Delta_d = 44$~MHz from the zero-velocity $5P_{1/2}, F=3 \rightarrow 5D_{3/2}, F=4$ resonance.

The tunable $\approx$~1280-nm coupler laser drives the transition to the $25F_{5/2}$ Rydberg state. While the coupler laser is scanned, the transmission of the locked probe is detected using a photo-diode [PD in Fig.~\ref{fig:setup}~(a)]. The PD current is amplified in a transimpedance amplifier, and the resultant EIT signal is recorded on an oscilloscope. Each experimental spectrum is an average of typically 300 scans. To calibrate the coupler-laser frequency scan, a portion of the 1280-nm laser power is sent through a Fabry-P\'erot (FP) etalon with a free spectral range of 374~MHz whose transmission resonances are recorded simultaneously with the spectroscopic measurements and used as frequency test points in post-processing. For higher-resolution coupler-laser frequency calibration, we also RF-modulate the beam sample in a fiber electro-optic modulator before passing it through the FP, generating a higher density set of frequency test points.

The probe and the coupler lasers are co-aligned in the measurement cell [$+z$ direction in Fig.~\ref{fig:setup}~(a)], while the dressing beam is counter-aligned to them ($-z$ direction). We use various dichroic mirrors [DM in Fig.~\ref{fig:setup}~(a), only one is shown] to align the lasers of the different wavelengths through the cell.
The counter-alignment of the probe and dressing beams helps with preventing leakage of dressing light onto the PD used to record the EIT spectra. Further, by routing the beams through polarization-maintaining fibers we ensure that all lasers have horizontal polarization [parallel to $x$ in Fig.~\ref{fig:setup}~(a)]. The beams are focused into the cell with full-widths-at-half-maxima (FWHM) of the intensity of $\approx$~80~$\mu$m for the probe and $\approx$~250~$\mu$m for the dressing beam and the coupler. Typical powers are $\approx$~10~$\mu$W, $\approx$~8~mW and $\approx$~15~mW for the probe, the dressing and the coupler beams, respectively. To raise the three-photon EIT signal, we first find and optimize the two-photon EIT signal on the $5S_{1/2} \rightarrow 5P_{1/2}\rightarrow5D_{3/2}$ cascade [see Fig.~\ref{fig:setup}~(b)] while scanning the dressing beam (with the coupler blocked). With the dressing beam properly aligned and locked, we un-block the coupler and scan it to search for the three-photon EIT signal. Once the three-photon EIT signal has been located, it is optimized further by fine-tuning the coupler alignment. 

To drive RF Rydberg-Rydberg transitions from $25F_{5/2}$, we use two different antennas to access different frequency ranges. The first is a log-periodic planar printed-circuit-board (PCB) antenna with 7~dBi gain for frequencies between 1 and 10~GHz, which emits an $x$-polarized field [see Fig.~\ref{fig:setup}~(a)]. This RF line starts with a signal generator with power $P_1$, followed by a coaxial cable (11~dB loss) connection to a 40-dB-gain amplifier and PCB antenna. The antenna is mounted above the cell at a distance of $d\approx15$~cm, which is marginally in the antenna far-field, and is oriented to co-align the polarization of the microwave with the laser polarizations in the cell. In this study, the PCB antenna is used to drive the $25F_{5/2} \leftrightarrow 25G_{7/2}$ and $25F_{5/2} \leftrightarrow 25H{9/2}$ transitions [field with frequency $f_1$ and power $P_1$ in Fig.~\ref{fig:setup}~(b)]. A second telescopic rabbit-ears antenna is fed directly by a second signal generator to apply the RF field driving the $25H_{9/2} \leftrightarrow 25I_{11/2}$ transition [field with frequency $f_2$ and power $P_2$ in Fig.~\ref{fig:setup}~(b)]. The rabbit-ears antenna, shown in Fig.~\ref{fig:setup}~(a), is folded such that it brackets the vapor cell in the $xz$-plane with its ears parallel along $z$ to generate linear RF polarization along $x$, and a small offset in $-y$ to avoid obstructing the alignment of the laser beams. The optimal leg length, which is on the order of $c /(4f_2)$, where $c$ is the speed of light, is kept at $L_2 \approx$~22~cm to maximize the RF field inside the cell.

\subsection{$25F_{5/2} \longleftrightarrow 25H_{9/2}$ transition}
\label{subsec:FH}

\begin{figure}[t!]
 \centering
  \includegraphics[width=0.45\textwidth]{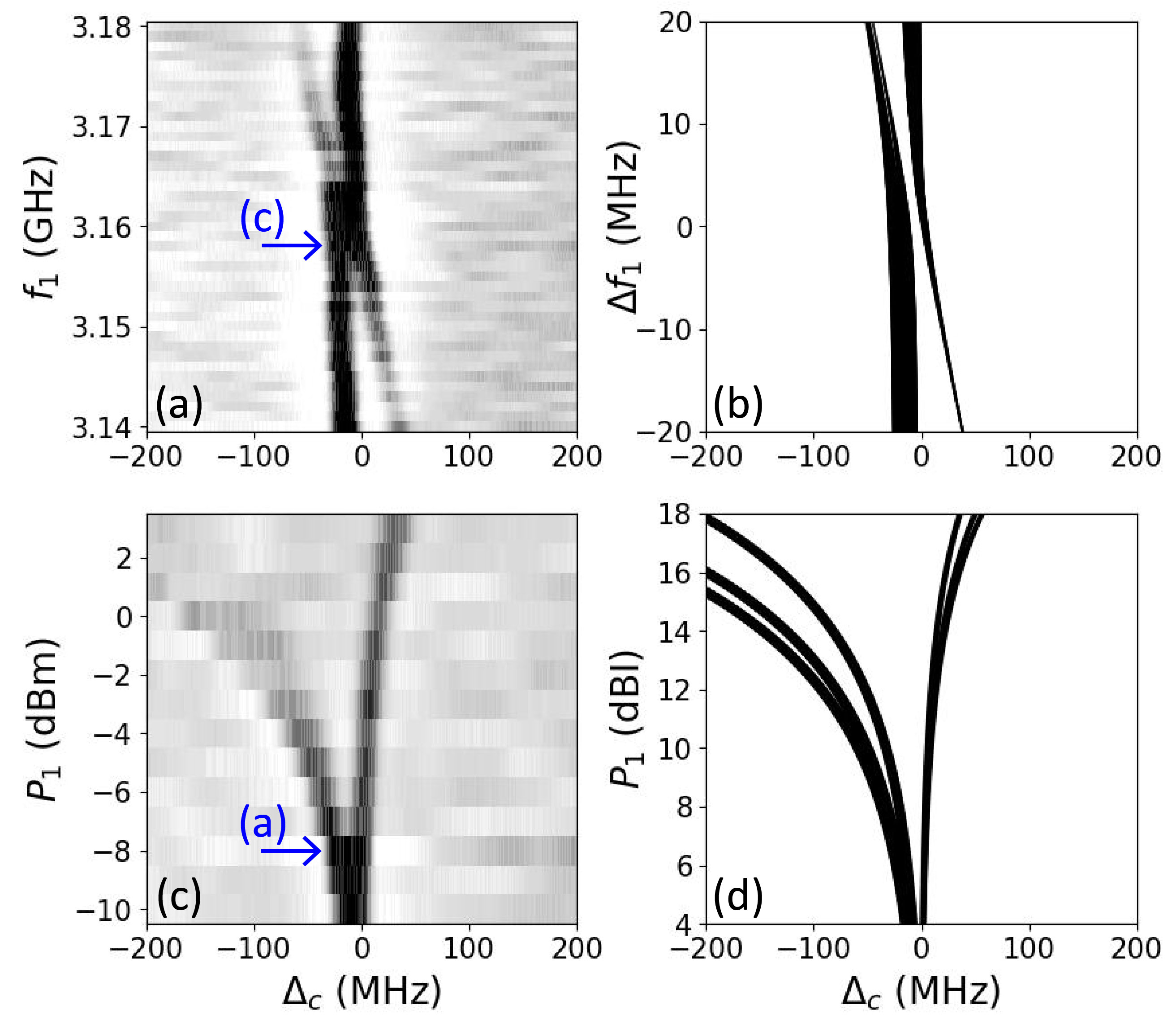}
  \caption{Spectroscopy of the $25F \longleftrightarrow 25H$ two-photon transition vs coupler-laser detuning. Experimental frequency and power scans are shown in (a) and (c) respectively, with the corresponding calculations in (b) and (d). In the experimental data in (a) and (c), the $y$-axes show absolute frequency, $f_1$, and power at the signal generator, $P_1$, respectively, and the signals are displayed on a linear gray scale. In the theory plots the areas of the symbols show approximate signal strength. In (b) the $y$-axis shows half of the two-photon detuning from the low-field resonance, and in (d) intensity in dBI (0~dBI = 1~W/m$^2$).} 
  \label{fig:FH}
\end{figure}

Experimental scans of $f_1$ and $P_1$ for the log-periodic antenna driving the two-photon $25F_{5/2}\leftrightarrow25H_{9/2}$ transition are shown in Figs.~\ref{fig:FH}~(a) and~(c) respectively (the rabbit-ears antenna is off). In (a) $P_1 = -8$~dBm at the signal generator, corresponding to about 21~dBm injection into the antenna, after taking cable losses and amplifier gain into account. With the 7~dBi antenna gain, we estimate an RF electric field at the atoms of about 41~V/m. Corresponding calculations of dressed-state energy levels for the optically accessed $m_J=1/2$, $3/2$ and $5/2$ manifolds are shown in Fig.~\ref{fig:FH}~(b) and~(d), with symbol area proportional to the $F$-state probability in the dressed states, which gives an approximate relative measure for EIT signal strength. In the calculations we use the absolute intensity unit dBI defined with reference to $I_0$=1~W/m$^2$, {\sl{i.e.}} the intensity $P($dBI$) = 10 \log_{10} [I / (1~$W/m$^2) ]$ with intensity $I$ in SI units.
The calculations include off-resonant AC shifts of all states in the RF field and utilize computed two-photon matrix elements between  $\vert 25F_{5/2}, m_J \rangle$ and $\vert 25H_{9/2}, m_J \rangle$.

Generally, we obtain very good agreement between the experimental data and the supporting computations, including measured two-photon Autler-Townes (AT) splittings and electric-field signal strengths. There are three AT pairs, one for each optically-coupled $m_J$. From Fig.~\ref{fig:FH}~(a) one can conclude that the two-photon resonance occurs at $\approx$~3.158~GHz, {\sl{i.e.}} the weak-field energy level difference is $h \times 6.316$~GHz, which agrees with the calculated energy-level difference to within 1~MHz. As $P_1$ is increased, the red-shifted AT peak in Fig.~\ref{fig:FH}~(c) exhibits larger shifts and broadening than the blue-shifted one, and approaches the noise floor at $P_1 \gtrsim$~1~dBm. The strong $m_J$-splitting of the lower AT branch, visible in (d) at an intensity of $P_1 > 12$~dBI, is marginally resolved in the experimental data in (c).  The asymmetry of the pattern in Figs.~\ref{fig:FH}~(c) and (d) results from the combination of AT splitting and AC shifts. By comparing measured and calculated spectra in Figs.~\ref{fig:FH}~(c) and (d) one obtains an approximate atomic power calibration of $P_1 (\text{dBI}) = P_1 (\text{dBm}) + $~14.5~dB. Taking into account the 1~dBm step-size in the experimental data in (c), the calibration uncertainty on this is $\pm$~1dB. Also, the RF electric-field amplitude at $P_1=-8~$dBm is $E_{RF1} = 58 \pm 7$~V/m, which is sufficiently precise for the present objective. The measured field exceeds the field estimated from the RF transmission line by about 40\%, which may be attributable to the fact that the vapor cell is only borderline in the far field.

\subsection{$25F_{5/2} \longleftrightarrow 25H_{9/2} \longleftrightarrow 25I_{11/2}$ transitions}
\label{subsec:FHI}

\begin{figure}[htb]
 \centering
  \includegraphics[width=0.45\textwidth]{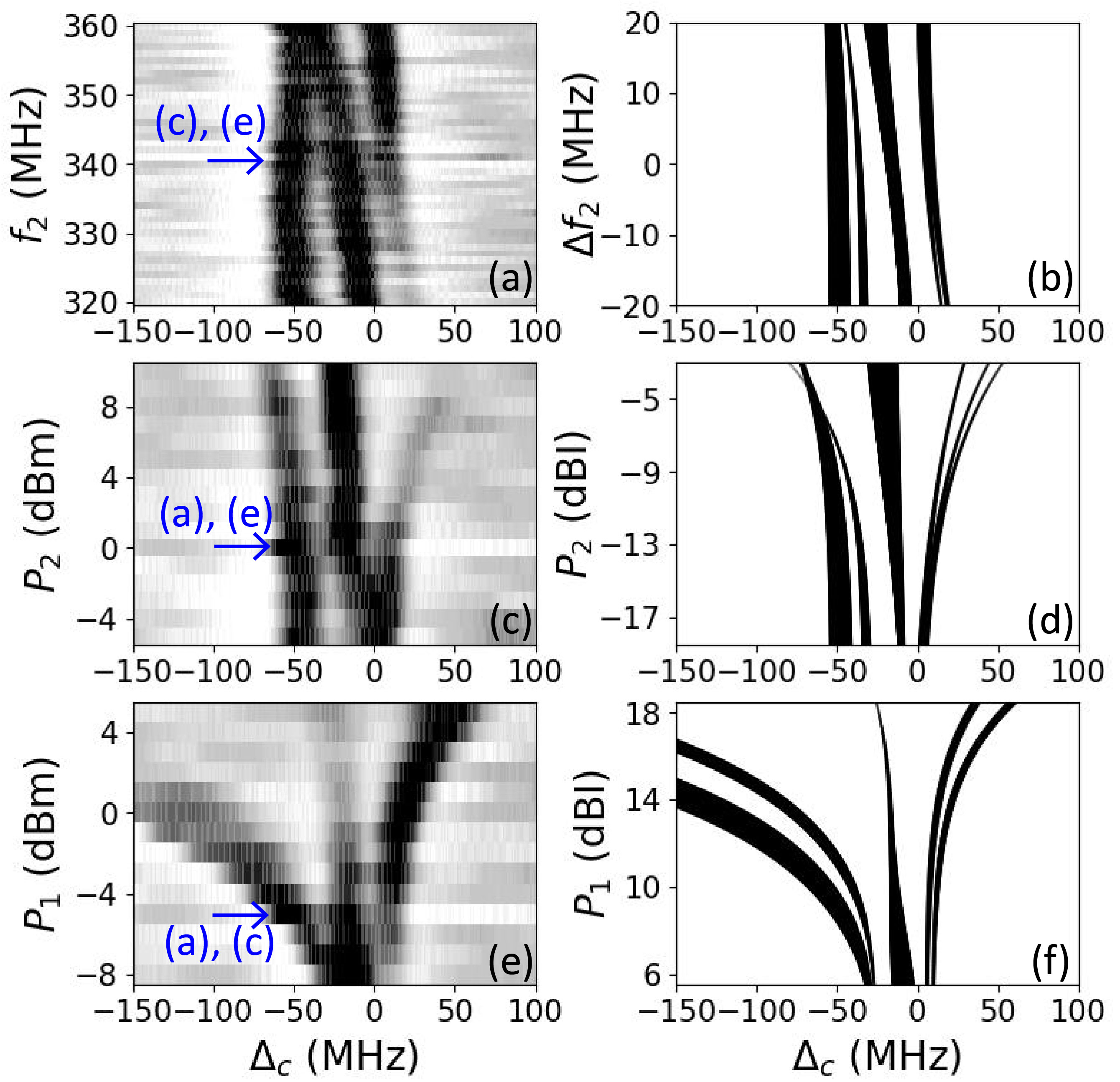}
  \caption{Spectroscopy of the $25F \longleftrightarrow 25H \longleftrightarrow 25I$ transition with both RF fields on vs coupler-laser detuning. Experimental frequency and power scans of the signal generator driving  the 25H$\longleftrightarrow$25I transition are shown in (a) and (c), respectively. A power scan of the signal generator driving  the $ 25F \longleftrightarrow 25H$ transition is shown in (e). The $y$-axes show absolute frequency, $f_1$, in (a), and powers at the respective signal generators in (c) and (e), and the signals are displayed on a linear gray scale.  Corresponding calculations are presented in (b), (d) and (f). In the theory plots, the areas of the symbols show approximate signal strength. The $y$-axis in (b) shows detuning from the 25H$\longleftrightarrow$25I resonance, and those in (d) and (f) RF field intensities in units dBI.} 
  \label{fig:FHI}
\end{figure}

We use the rabbit-ears antenna to drive the $25H_{9/2}\leftrightarrow25I_{11/2}$ transition. In this case, there are three RF-coupled Rydberg states (per $|m_J|$), leading to the three dressed-state peaks in the experimental spectra in Fig.~\ref{fig:FHI}. The $|m_J|$-splittings are only resolved in the calculations.  The calculations include off-resonant AC shifts of all states caused by both RF fields, and we use two-photon matrix elements for the $\vert 25F_{5/2}, m_J \rangle$ $\leftrightarrow$ $\vert 25H_{9/2}, m_J \rangle$ and one-photon matrix elements for the $\vert 25H_{9/2}, m_J \rangle$ $\leftrightarrow$ $\vert 25I_{11/2}, m_J \rangle$ transitions. As before, only the optically accessed $m_J=1/2$, $3/2$ and $5/2$ manifolds are relevant.  While $m_J$-splittings occur in the calculations in Figs.~\ref{fig:FHI}~(b), (d) and (f), these are not resolved in the present experimental data, leaving three broadened peaks in the data in (a), (c) and (e). Regardless, overall good agreement between the experimental data and the theory is observed.  

For the scan of $f_2$ in Fig.~\ref{fig:FHI}~(a), it is $P_1= -5$~dBm and $f_1=3.158$~GHz, the resonance frequency found in Fig.~\ref{fig:FH}, and the power into the rabbit-ears antenna is set at $P_2 =$~0~dBm. The $25H_{9/2}\leftrightarrow25I_{11/2}$ resonance is observed at $\approx$~340~MHz, which agrees with our calculated value of $339.6$~MHz within uncertainties. The value $f_2=340$~MHz is chosen for the $P_1$- and $P_2$-scans in Figs.~\ref{fig:FHI}~(c) and (e).

Investigating the power dependence, we observe that as $P_2$ is increased in Fig.~\ref{fig:FHI}~(c), while holding $P_1$ at $-5$~dBm, the rightmost peak blue-shifts and disappears at $P_2 \gtrsim $~8~dBm. In the corresponding calculations in Fig.~\ref{fig:FHI}~(d), the rightmost peak $m_J$-splits into three lines, while their approximate signal strengths (proportional to symbol area in the theory plots) significantly diminish, in agreement with the experimental data in Fig.~\ref{fig:FHI}~(c). At the same time, the leftmost and the central lines remain relatively strong throughout the tested $P_2$-range and experience small red shifts. As $P_2$ is increased, in experiment [Fig.~\ref{fig:FHI}~(c)] and calculation [Fig.~\ref{fig:FHI}~(d)] the oscillator strength redistributes into the central line. 
Comparing Figs.~\ref{fig:FHI}~(c) and~(d), we calibrate $P_2$ and find $P_2 (\text{dBI}) = P_2 (\text{dBm}) - $~13.5~dB $\pm$~1~dB. The main source of uncertainty is the step size in $P_2$ of 1~dBm. Again, in the present context an approximate atomic calibration is sufficient. At a power of $P_2= 0$~dBm the electric-field amplitude $E_{RF2} = 5.8 \pm 0.7$~V/m. For this field, a distance of 3~cm between the legs of the rabbit-ears antenna, and for impedance-matched coupling, the RF power feeding the antenna would be about -10~dBm. Factoring in transmission-line losses and some (likely) impedance mismatch, this is reasonably close to the actual signal generator power of $P_2=0$~dBm.

Finally, we explore the effect of varying $P_1$ on the three-peak EIT spectra in  Fig.~\ref{fig:FHI}~(e). Here $f_2 =$~340~MHz and $P_2 =$~0~dBm. As can be seen in Fig.~\ref{fig:FHI}~(e), the observed shifts as well as the signal-strength distribution change quite dramatically; this is because the $F$ to $H$ transition is a 2-photon transition. First, the rightmost peak exhibits positive shifts, similar to Fig.~\ref{fig:FHI}~(c). However, opposite to Fig.~\ref{fig:FHI}~(c), it becomes the strongest peak. This peak would be expected to $m_J$-split [see Fig.~\ref{fig:FHI}~(f)]. Due to the intrinsic EIT linewidth and likely inhomogeneities of the RF intensities within the cell, in the experiment we observe line broadening instead of splitting [see Fig.~\ref{fig:FHI}~(e)]. Second, the central peak, while slightly red-shifting, almost disappears at $P_1 >$~3~dBm, in accordance with the calculations in Fig.~\ref{fig:FHI}~(f). Finally, we observe that the leftmost line massively red-shifts (by $\approx$~150~MHz at $P_1 \sim $~0~dBm). Here, the $m_J$-splitting, seen in the calculation in Fig.~\ref{fig:FHI}~(f), is borderline resolved in the experiment.

\section{DC-field sensing}
\label{sec:dcsensing}

\begin{figure*}[t!]
 \centering
  \includegraphics[width=0.99\textwidth]{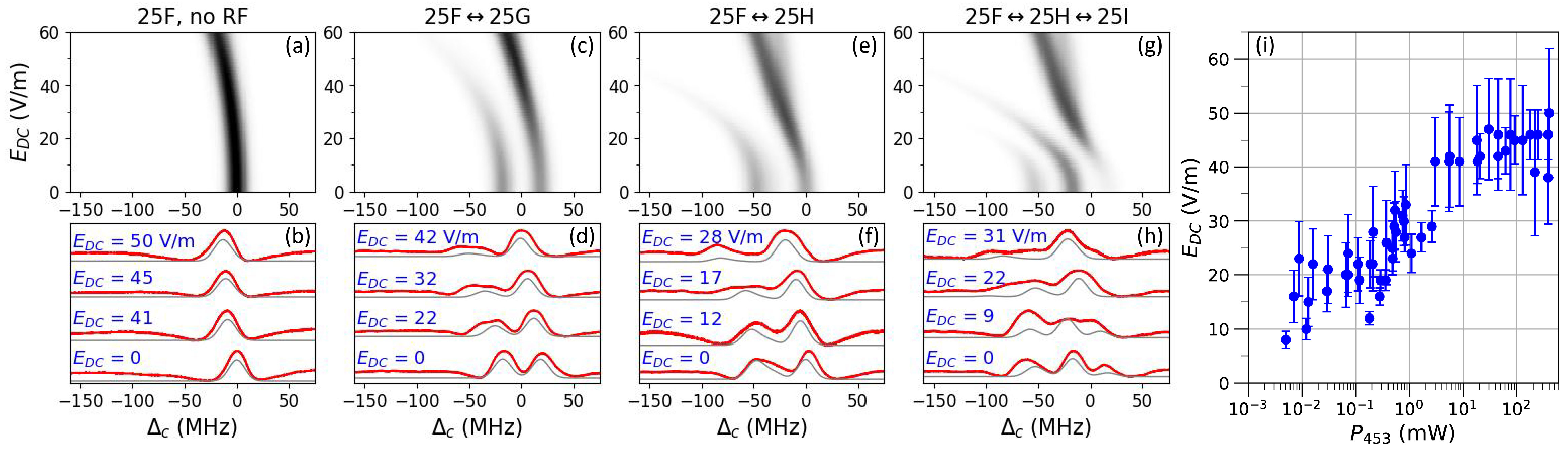}
  \caption{Panels (a), (c), (e), and (g):
  Calculated EIT spectra vs coupler-laser detuning, $\Delta_c$, and DC electric field, $E_{DC}$, with the indicted RF transitions applied. Corresponding experimental EIT spectra in selected light-induced DC electric fields are shown in panels (b), (d), (f), and (h) (bold red curves), together with corresponding spectra at fixed $E_{DC}$ from the plots above that best match the experimental data (gray curves). The best-fit $E_{DC}$-values extracted from the match of each measured spectrum to the calculated spectra are indicated. Best-fit $E_{DC}$-values versus 453-nm photo-illumination laser power, $P_{453}$, are shown in panel~(i).}
  \label{fig:DC}
\end{figure*}

In this Section we utilize Rydberg states with $\ell \leq$~6 prepared as described in the previous Section for DC-electric-field sensing. The method of creating DC fields by photo-illumination of the vapor-cell wall is first reviewed, followed by experimental measurements of the DC fields. The atomic-calibration-based estimates for $E_{RF1}$ and $E_{RF2}$ are employed to model the experimental data.

\subsection{Creating DC fields in vapor cell by wall photo-illumination}
\label{subsec:photofields}

Applying DC electric fields to atomic vapors confined in glass cells has turned out to be a considerable challenge~\cite{mohapatra2007, Sedlacek2012,Holloway2014}, as these devices exhibit a robust isolation against external static fields. Random, weak electric fields (sometimes referred to as microfields)~\cite{hooper, potekhin,demura} may still be present and require an experimental characterization~\cite{hollowayjapl2017}. Charges generating the microfields may arise from effects including photo-ionization of atoms, and collisions involving atoms in Rydberg~\cite{gallagher, Beterov_2009} or low-lying intermediate states (such as $5P_{1/2}$ and $5D_{3/2}$)~\cite{Cheret_1982, barbier3, Barbier1, Barbier2}. 

To demonstrate DC electric field measurement using high-$\ell$ Rydberg states, in the present work we require a simple method to generate DC test fields in the cell. While electrodes inside the cell or the cell walls~\cite{barredo2013, Grimmel_2015,andersonapl2018,ma2022} can be used for that, no such electrodes are present in our experiments. Therefore, we use an alternative method in which free charges are generated on the cell walls by laser illumination and photo-electric effect~\cite{Rousseau1968}. In~\cite{hankin2014,jau2020}, this approach was utilized to create patch regions with local electric fields on glass surfaces. It has also been shown that the light-induced DC fields can be quite homogeneous and reach up to $\sim$~60~V/m within the vapor cell~\cite{ma20}.

In our work, a 453-nm laser beam of $\approx$~1~cm diameter, that forms a small ($\approx$~20$^{\circ}$) angle with respect to the cell axis, is aligned in a way that it illuminates a cell-wall section close to one of the legs of the rabbit-ears antenna along the entire length of the cell. This generates an approximately homogeneous DC electric field, $E_{DC}$, within the EIT laser beams passing through the cell. The field $E_{DC}$ is approximately parallel to all laser and RF electric fields in our setup, which all point along $x$ in Fig.~\ref{fig:setup}~(a). The dependence of the $E_{DC}$-magnitude on 453-nm laser power is highly nonlinear~\cite{ma20}. Hence, in addition to showing that high-$\ell$ Rydberg-EIT is suitable for measurement of DC electric fields in vapor cells, in effect we also determine what 453-nm laser power results in what level of $E_{DC}$ in our experiment.

\subsection{Overview of results}
\label{subsec:results}

In Figs.~\ref{fig:DC}~(a), (c), (e) and~(g), we plot calculated EIT maps versus coupler-laser detuning and $E_{DC}$. The calculation for the RF-dressed Rydberg levels is similar to those in Secs.~\ref{subsec:FH} and~\ref{subsec:FHI}, with the addition of the DC Stark shifts using the DC polarizabilities from Table~\ref{tab:table1}. This yields RF-dressed Rydberg-level energies, $\Delta W(k,m_J)$, and corresponding $|25F, m_J \rangle$-amplitudes, $c(k,m_J)$, where $k$ is a dressed-state label counting from 1 to up to 3 for each $m_J$ (which is conserved for all-parallel electric fields). Only 
levels with $|m_J|=1/2$, $3/2$ or $5/2$ are coupled. We then use an empirical finding according to which the EIT line strengths scale with $| \Omega_C \Omega_P|^2$, where $\Omega_P$ is the probe Rabi frequency on the transition $\vert 5S_{1/2}, F=3, m_F \rangle \leftrightarrow \vert 5P_{1/2}, F'=3, m_{F'}=m_F \rangle$, and $\Omega_C$ is the effective two-photon coupler Rabi frequency on the transition $\vert 5P_{1/2}, F'=3, m_F \rangle \leftrightarrow \vert 25F_{5/2}, F''', m_F \rangle$ transition. $\Omega_C$ is coherently summed over all intermediate, detuned states $\vert 5D_{3/2}, F'', m_F \rangle$, where the summing index is $F''$. Performing all sums over $m_F$ and the unresolved Rydberg hyperfine structure, $F'''$,  one finds EIT line strengths proportional to $S(k,|m_J|) = |c(k,|m_J|)|^2 w(|m_J|)$, where the Rydberg physics is contained in the determination of the  $|25F, |m_J| \rangle$-amplitudes in the RF-dressed states, $c(k,|m_J|)$, and the EIT readout physics in the $|m_J|$-weights, $w(|m_J|)$. The magnitude-squares of $c(k,|m_J|)$ correspond with the areas of the symbols in Figs.~\ref{fig:FH} and~\ref{fig:FHI}, whereas in Fig.~\ref{fig:DC} we use the $S(k,|m_J|)$.  The weights $w(|m_J|)$ depend somewhat on the intermediate-state detuning $\Delta_d$ in Fig.~\ref{fig:setup}~(b) and are, in our case, about $w(|m_J|)=0.43$, 0.29, and 0.28 for $|m_J|=1/2$, $3/2$, and $5/2$, respectively. EIT is observed when the coupler detuning $\Delta_c$ equals to one of the dressed-state energies, $\Delta W(k,|m_J|)$. To obtain the model spectra displayed in the top row in Fig.~\ref{fig:DC}, we convolute the EIT lines with identical Gaussian line profiles with FWHM of 18~MHz, the approximate width of the $25F$ EIT line without RF dressing fields, and sum over all $(k,|m_J|)$.

In Figs.~\ref{fig:DC}~(b), (d), (f) and~(h), exemplary experimental EIT spectra for the corresponding cases are shown, together with the best-matching spectra from the calculated maps above. For clarity, the calculated spectra are slightly offset from the experimental data. Measured and computed spectra are matched empirically, which is sufficient at the present level. In Fig.~\ref{fig:DC}~(i) we plot the matched values of $E_{DC}$ from all other panels of the figure versus 453-nm photo-illumination power, $P_{453}$. There, we approximately calibrate the DC electric field as a function of illumination power using high-$\ell$ Rydberg EIT spectroscopy for our experimental setup. In the remainder of the paper, we discuss these results in detail.

\subsection{Detailed analysis of results}
\label{subsec:resultsdetails}

In Figs.~\ref{fig:DC}~(a) and~(b) it is apparent that the $25F_{5/2}$-state, which has the lowest polarizabilities among the states tested here, only exhibits significant Stark shifts, exceeding about 10~MHz, for $E_{DC} \gtrsim 40$~V/m, which already is near the largest fields that we can generate with the photo-illumination method used.  As $E_{DC}$ is increased further to 50~V/m (by increasing $P_{453}$), the experimental spectra merely shift by an additional 3-5~MHz and exhibit a slight broadening. Increasing $n$ aids in achieving a stronger response of $nF_{5/2}$ states. For example, the $57F_{5/2}$ EIT peak was seen to disappear entirely at our maximum $P_{453}$ (not shown). In the following, we employ RF-dressing at $n=25$ to access higher-$\ell$ Rydberg states at the same $n$ to elicit a stronger DC electric-field response, afforded by the higher polarizabilities of those states (see Table.~\ref{tab:table1}).

Increasing $\ell$ by driving the $25F_{5/2} \leftrightarrow 25G_{7/2}$ transition with the log-periodic antenna ($f_1 =$~5.234~GHz, $P_1$ = -20~dBm), moderately enhances the response, as seen in Figs.~\ref{fig:DC}~(c) and~(d). The two AT peaks start to notably change at $E_{DC} \gtrsim $~20~V/m. The left AT peak shifts and broadens considerably more than the right one. At fields $E_{DC} \gtrsim$~40~V/m, the left AT peak diminishes greatly in signal strength and eventually becomes indiscernible while becoming more responsive to changes in $E_{DC}$. This behavior reflects the fact that the dressed state associated with the left AT peak turns into the stronger-shifting $25G_{7/2}$ state, which has a larger polarizability than the $25F_{5/2}$ state. As the $25G_{7/2}$-character of the lower AT peak increases, its EIT signal strength diminishes due to the diminishing optical coupling to the $5D_{3/2}$ state [see Fig.~\ref{fig:setup}~(b)]. Conversely, with increasing $E_{DC}$ the right AT peak acquires an increasing amount of $25F_{5/2}$ character, which reduces its response to changes in $E_{DC}$ while its EIT signal strength increases.

Increasing $\ell$ yet more by driving the $25F_{5/2} \leftrightarrow 25H_{9/2}$ two-photon microwave transition (see Sec.~\ref{subsec:FH}) allows for further $E_{DC}$-response enhancement, as seen in Figs.~\ref{fig:DC}~(e) and~(f). Here $f_1$ and $P_1$ for the log-periodic antenna are 3.158~GHz and -5~dBm, respectively. Already at $E_{DC} \gtrsim $~10~V/m the two AT peaks significantly change. The left AT peak in Figs.~\ref{fig:DC}~(e) and~(f) shifts and broadens considerably more than in Figs.~\ref{fig:DC}~(c) and~(d), and at $E_{DC} \gtrsim$~30~V/m its strength drops to hardly noticeable. These features reflect the even larger polarizability of the  $25H_{9/2}$ state, which lowers the detectable $E_{DC}$. The excess broadening [Fig.~\ref{fig:DC}~(e)] and splitting [Fig.~\ref{fig:DC}~(f)] of the  $25F_{5/2}$-like upper AT peak at fields $E_{DC} \gtrsim $~30~V/m results from an increase of the $m_J$-splitting caused by the AC Stark effect of the $25F_{5/2}$-state in the rather strong two-photon RF drive field, $E_{RF1} \approx 80$~V/m.

Finally, we utilize both the PCB and the rabbit-ears antennas to assess the response of the three-peak EIT on the $25F_{5/2} \leftrightarrow 25H_{9/2} \leftrightarrow 25I_{11/2}$ ladder to the DC fields in Fig.~\ref{fig:DC}~(h). 
Here, $f_1$ and $P_1$ are same as in Fig.~\ref{fig:DC}~(f), while $f_2$ = 345~MHz and $P_2$ = 4~dBm.
In this most $E_{DC}$-field-sensitive case, we have utilized finer steps in $P_{453}$ (and, hence, in $E_{DC}$) 
to resolve the spectral changes in fields $E_{DC} \lesssim $~10~V/m. First, the rightmost peak appears to red-shift and to merge into the central peak, and the signal strength redistributes between the peaks. At $E_{DC} \gtrsim$~20~V/m first the most red-shifting line acquires increasing $25I_{11/2}$-character and greatly diminishes in signal strength, followed by an increasingly $25H_{9/2}$-like state departing to the red-shifted side. At larger fields, the $25F_{5/2}$-like peak dominates and eventually becomes the only visible peak left, and the spectrum looks similar to Figs.~\ref{fig:DC}~(e) and~(f). 

In all four cases in Fig.~\ref{fig:DC}, the theoretical predictions qualitatively agree with the experimental data. Fig.~\ref{fig:DC} demonstrates the step-wise increase in DC electric-field sensitivity with increasing $\ell$, probed via three-photon optical Rydberg EIT. The minor discrepancies between measurements and calculations are attributed to imperfections in the optical and RF polarizations, inhomogeneities in both the RF and DC fields, and environmental magnetic fields (which were not shielded). Especially at small $E_{DC}$, the latter may cause $m_J$-mixing and Zeeman shifts of the high-$\ell$ states on the order of $\pm$5~MHz. In view of Table~\ref{tab:table1}, the $m_J$-mixing may alter the spectra due to the large tensor polarizabilities. 

At $E_{DC}\sim 0$ and for RF parameters that lead to efficient mixing of the coupled levels, the polarizabilities of the mixed levels are approximately given by the averages over the polarizabilities of the coupled states.  In Fig.~\ref{fig:DC}, these averages are to be taken over the levels printed over the tops of the four data sets.
Assuming that the $|m_J|$-range is restricted to $|m_J| \leq 5/2$, {\sl{i.e.}} the states that are optically coupled for all $x$-polarized fields, the averages are 0.010, 0.027, 0.068 and 0.145~MHz/(V/m)$^2$ for Figs.~\ref{fig:DC}~(a), (c), (e) and~(g), respectively. At small $E_{DC}$, all dressed levels in a given plot should Stark-shift with roughly equal curvatures, and the curvature ratios between the four sets of plots in Figs.~\ref{fig:DC}~(a), (c), (e) and~(g) should roughly accord with the stated numbers. This simplified analysis explains the trends that are observed in Fig.~\ref{fig:DC} for $E_{DC} \lesssim 15$~V/m, {\sl{i.e.}} in regions where the RF coupling amplitudes exceed the Stark shifts for all four cases.

In Fig.~\ref{fig:DC}~(i) we plot the $E_{DC}$-values obtained by matching observed and calculated spectra from  Figs.~\ref{fig:DC}~(a) through~(h) versus the 453-nm photo-illumination power, $P_{453}$, used to generate the DC field.
The plot reflects fairly large, conservative estimates for the $E_{DC}$-uncertainty. The latter are obtained from matching the experimental spectra to different calculated spectra empirically and are usually between 10-20\% of respective $E_{DC}$. For $P_{453} \lesssim 10$~mW the field increases approximately linearly in $\log(P_{453})$ and saturates at higher powers at values near 50~V/m, in general agreement with earlier findings~\cite{ma20}. 
The observed highly nonlinear behavior reflects a complex charging or discharging equilibrium within the vapor cell. This topic is beyond the scope of our present work and may be explored further in future studies.

\section{Discussion and Outlook}
\label{sec:disc}

We have prepared Rydberg states with $\ell \leq$~6 in a room-temperature vapor cell using a combination of three-photon optical EIT and up to three-photon RF dressing. The RF dressing fields were calibrated by Rydberg-EIT spectroscopy. The resultant well-characterized dressed Rydberg states were then used to measure and calibrate DC electric fields generated by photo-electric effect on the cell walls. The utility of the increased electric polarizabilities of the high-$\ell$ Rydberg states in measuring weak electric fields was demonstrated. The underlying physics was modeled by accounting for all AC and DC shifts and resonant couplings of the dressed Rydberg-atom system, as well as the specifics of the utilized three-optical-photon Rydberg-EIT probe.

Possible improvements in the setup to extend the electric field sensitivity could include stray-magnetic-field  shielding, reduction of the EIT laser linewidths, and reduction of RF field inhomogeneities and unwanted polarization variations~\cite{shaffer2018,jing2020,prajapatiapl2021,hollowayapl2022,dixon2023, legaie2023}. Solving the Lindblad equation for our case of collinear three-optical-photon Rydberg EIT in a Rb vapor cell, we find that EIT linewidths $<5$~MHz should be possible. At the low $n$-value used in our work, optical electric-dipole transition matrix elements are relatively large (scaling $\sim n^{-3/2}$) and RF-transition ones relatively small (scaling $\sim n^2$). These scalings favor economical setups, as laser power is more expensive than RF power at sub-5-GHz frequencies.   
We therefore anticipate applications in high-precision measurements on relatively low-$n$ and high-$\ell$ Rydberg states~\cite{lee,berl2020,moore2020} in room-temperature vapor cells for tests of \textit{ab initio} atomic-physics calculations~\cite{marinescu1994, safronova2004,safronova2011,gaugeeffects}, the exploration of pathways towards excitation of circular Rydberg atoms~\cite{anderson2013, zhelyazkova2016, larrouy2020, cardman2020, wu2023}, and other applications in fundamental-physics research~\cite{safronovarmp}. 

Along the lines of applications mentioned in the introduction, the demonstrated methods could be of interest for sensing electric fields in low-pressure discharge plasmas~\cite{harry2013,eden2013,Nijdam_2022}. Combining Rydberg-EIT in low-pressure buffer-gas vapor cells, which was recently observed~\cite{thaicharoen2023}, with plasma generators, one may anticipate future plasma-physics research in discharge plasmas with Rydberg-EIT as an electric-field probe. This may include dusty plasmas~\cite{shukla2001,shuklarmp2009}, which are not only relevant to applications~\cite{Boufendi_2011, Ratynskaia2022} but also in astrophysics~\cite{mendis1994}, nonlinear dynamics~\cite{morfill2009} and condensed matter~\cite{thomas1994}. Rydberg-EIT as a non-invasive electric-field probe with high spatial resolution~\cite{anderson2017} employing high-$\ell$ states could become useful in future dusty-plasma research.

\section*{Acknowledgments}
\label{sec:acknowledgments}
We would like to thank Bineet Dash, Nithiwadee Thaicharoen, Michael Viray and Jamie MacLennan for useful discussions. This work was supported by the U.S. Department of Energy, Office of Science, Office of Fusion Energy Sciences under award number DE-SC0023090, NSF Grant No. PHY-2110049, and Rydberg Technologies, Inc.. A.D. and R.C. acknowledge support from the respective Rackham Predoctoral Fellowships at the University of Michigan. D.A.A and G. R. have an interest in Rydberg Technologies, Inc.

\bibliography{bibliography.bib}

\end{document}